\documentclass[12pt]{article}
\usepackage{a41}
\usepackage{epsfig}
\usepackage{axodraw}
\usepackage{eucal}

\newcommand\xt{\tilde{x}}
\newcommand\ep{\varepsilon}

\begin{document}
\setlength{\baselineskip}{0.515cm}
\sloppy
\thispagestyle{empty}
\begin{flushleft}
DESY 01--032 \hfill
{\tt hep-ph/0104099}\\
April  2001  \\
\end{flushleft}

\setcounter{page}{0}

\mbox{}
\vspace*{\fill}
\begin{center}{
\LARGE\bf On the Anomalous Dimension of}

\vspace{2mm}
{\LARGE\bf
 the Transversity Distribution
{\boldmath $h_1(x,Q^2)$}}$^{\footnotemark}
$\footnotetext{Work supported in part by EU contract FMRX-CT98-0194
(DG 12 - MIHT)}

\vspace{5em}
\large
Johannes Bl\"umlein$^{{\footnotemark}}
$\footnotetext{Electronic address: {\tt johannes.bluemlein@desy.de}}
\\
\vspace{5em}
\normalsize
{\it   DESY--Zeuthen}\\
{\it Platanenallee 6, D--15735 Zeuthen, Germany}\\

\vspace*{\fill}
\end{center}
\begin{abstract}
\noindent
We calculate the leading order anomalous dimension of the transversity 
structure function directly using three different methods, the local
light--cone expansion in the forward case, the non--forward case, and 
the short--distance expansion of the forward Compton amplitude. Our 
results agree with the original calculation by Artru and Mekhfi 
[Z. Phys. {\bf 45} (1990) 669],  which has been doubted recently.
We also comment on the next-to-leading order anomalous dimension.
\end{abstract}

\vspace{1mm}
\noindent
\begin{center}
PACS numbers~:~11.10.H1, 11.40.-q, 11.55.Ds, 12.38.Bx, 13.85.Fb, 13.88.+e

\end{center}

\vspace*{\fill}
\newpage
\section{Introduction}

\vspace{1mm}
\noindent
The transversity distribution $h_1(x,Q^2)$ \cite{RS,JJ} is one of the
central inclusive quantities which emerge in the transverse polarized 
Drell--Yan
process and deep inelastic scattering off targets with a spin $J \geq 1$.
This function plays also a crucial rule in proving sum rules for other
inclusive functions in deep inelastic scattering off polarized 
targets~\cite{BKBT}. The transversity distribution is a flavor 
non--singlet function and receives contributions starting with twist--2. 
The experimental measurement of $h_1(x,Q^2)$ is currently planned at 
different facilities~\cite{EXP} and the scaling violations of this 
quantity may ultimately be compared with the predictions of perturbative 
Quantum Chromodynamics and thus serve as a novel test of this theory.

The leading order (LO) splitting function $P_{h_1}(z)$ for the 
distribution $h_1(x,Q^2)$ has first been calculated by Artru and 
Mekhfi~\cite{AM} using time-ordered or `old-fashioned' perturbation 
theory~\cite{TOPT} for the range of momentum fractions $z < 1$ 
determining the end--point contribution as in the case of the 
longitudinally unpolarized or polarized flavor non--singlet distribution
$P_{NS}(z)$~\cite{FERMI}, where the conservation of fermion--number 
allows this.
For the transversity distribution such an integral relation does not
exist a priori. Therefore one may doubt this procedure, as has been done 
in Ref.~\cite{MKS} recently, where a different result was obtained
for the splitting function $P_{h_1}(z)$ in a direct calculation using
a method by Ioffe and Khodjamirian~\cite{IK}. As a result the authors
of \cite{MKS} conclude that either the splitting function $P_{h_1}(z)$ 
was wrongly calculated in the past or the method of Ref.~\cite{IK} is
inapplicable. 

It is the aim of the present paper to clarify this problem, which is
of importance for the correct understanding of the evolution of the
transversity distribution $h_1(x,Q^2)$. We will calculate the anomalous
dimension of the distribution $h_1(x,Q^2)$ by three different methods.
Moreover, it is interesting to know, whether one can determine anomalous 
dimensions using the method
\cite{IK}, which may be of special importance for time--like higher
order calculations, i.e. in a situations where the light--cone expansion
is {\it inapplicable}.
  A related question as raised in \cite{MKS} on the
correct end--point behaviour indeed occurs also in next-to-leading
order, since the splitting function there also was evaluated under some
assumptions~\cite{NLO}, due to the techniques being applied, 
see~\cite{CUT,CFP}.  

The paper is organized as follows.
In section~2 we recalculate the forward anomalous dimension of 
$h_1(x,Q^2)$ in leading order using the local light--cone expansion, by
calculating the non--forward anomalous dimension, from which the forward 
limit is derived, and the short--distance expansion of a related Compton
amplitude~\cite{IK}. In section~3 we comment on the next-to-leading
order anomalous dimension of $h_1(x,Q^2)$ and section~4 contains the
conclusions.
\section{LO Anomalous Dimension of {\boldmath $h_1(x,Q^2)$}}

\vspace{1mm}
\subsection{Forward-Scattering Anomalous Dimension}

\vspace{1mm}
\noindent
We first calculate the anomalous dimension using the local light cone
expansion in the forward case, $p_1 = p_2 = p$. The contributing diagrams
are depicted in Figure~1. The Feynman rules for the vector--valued
operator insertions for two quark and two quark and a gluon line read
\begin{eqnarray}
\label{eq1}
O_n^{(0) \mu}. \Delta &=& \sigma^{\mu\nu} \Delta_{\nu} (p.\Delta)^n \\
\label{eq2}
O_n^{(1) \mu \lambda}. \Delta &=& g t^a \sigma^{\mu\nu} \Delta_{\nu}
\sum_{j=1}^n (p'.\Delta)^{j-1} \Delta^{\lambda} (p.\Delta)^{n-j}~.
\end{eqnarray}
They are denoted by the symbol {\boldmath $\otimes$} in Figure~1. Here,
$p'$ is the second quark momentum at the quark-quark-gluon insertion
Figure~1b,c,
$\sigma_{\mu\nu} = (i/2) [\gamma_\mu \gamma_\nu - \gamma_\nu
\gamma_\mu]$, $\Delta_\rho$ is a light--like vector with $\Delta . \Delta
=0$,
$t^a$ denotes the $SU(3)_c$ generators, $g$ is the strong coupling 
constant, and $\lambda$ the vector index of the additional gluon.

The anomalous dimension $\gamma_n(g(\mu))$ is obtained by
\begin{eqnarray}
\label{eq3}
\gamma_n(g(\mu)) = \mu^2 \frac{\partial}{\partial \mu^2}~\ln Z_n(\mu)~,
\end{eqnarray}
with
\begin{eqnarray}
\label{eq4}
O_n(\mu) = Z_n^{-1}(\mu)~O_n^{\rm bare}~.
\end{eqnarray}
In leading order the anomalous dimension is a gauge invariant quantity
\footnote{In the $\overline{\rm MS}$ scheme this holds to all orders.}.
We will work, for definiteness, in the Feynman--gauge. The contributions
to the $Z$--factor in the forward case will be denoted by $I_j$ with
$j=a, ..., e$, cf. Figure~1.

\vspace*{-2.5cm}
\begin{center}
\begin{picture}(270,100)(0,0)
\setlength{\unitlength}{0.05mm}
\SetWidth{1.2}
\ArrowLine(-100,-100)(-80,-50)
\ArrowLine(-80,-50)(-60,0)
\ArrowLine(-60,0)(-40,-50)
\ArrowLine(-40,-50)(-20,-100)
\Gluon(-80,-50)(-40,-50){5}{7}
\BCirc(-60,0){7}
\ArrowLine(-10,-100)(10,-50)
\ArrowLine(10,-50)(30,0)
\ArrowLine(30,0)(70,-100)
\GlueArc(20,-25)(26.925824,246.422,66.422){5}{12}
\BCirc(30,0){7}
\ArrowLine(80,-100)(120,0)
\ArrowLine(120,0)(140,-50)
\ArrowLine(140,-50)(160,-100)
\GlueArc(130,-25)(26.925824,113.578,-66.422){5}{12}
\BCirc(120,0){7}
\ArrowLine(190,-100)(203.3333333,-66.66666666)
\ArrowLine(203.3333333,-66.66666666)(216.6666666,-33.3333333)
\ArrowLine(216.6666666,-33.33333333)(230,0)
\ArrowLine(230,0)(270,-100)
\GlueArc(210,-50)(17.95055,66.422,-113.578){5}{9}
\BCirc(230,0){7}
\ArrowLine(280,-100)(320,0)
\ArrowLine(320,0)(333.3333333,-33.33333333)
\ArrowLine(333.3333333,-33.33333333)(346.66666666,-66.66666666)
\ArrowLine(346.6666666,-66.66666666)(360.0,-100)
\GlueArc(340,-50)(17.95055,113.578,293.578){5}{9}
\BCirc(320,0){7}
\setlength{\unitlength}{1pt}
\Text(-100,-115)[]{$p_1$}
\Text(-20,-115)[]{$p_2$}
\Text(-60,-115)[]{$a$}
\Text( 30,-115)[]{$b$}
\Text(120,-115)[]{$c$}
\Text(230,-115)[]{$d$}
\Text(320,-115)[]{$e$}
\Text(180,-50)[]{$1/2~\biggl[$}
\Text(360,-50)[]{$\biggr]$}
\Text(275,-50)[]{$+$}
\Text(-60,0)[]{$\times$}
\Text( 30,0)[]{$\times$}
\Text(120.5,0)[]{$\times$}
\Text(231,0)[]{$\times$}
\Text(321,0)[]{$\times$}
\end{picture}
\end{center}

\vspace*{4cm} \noindent
\small
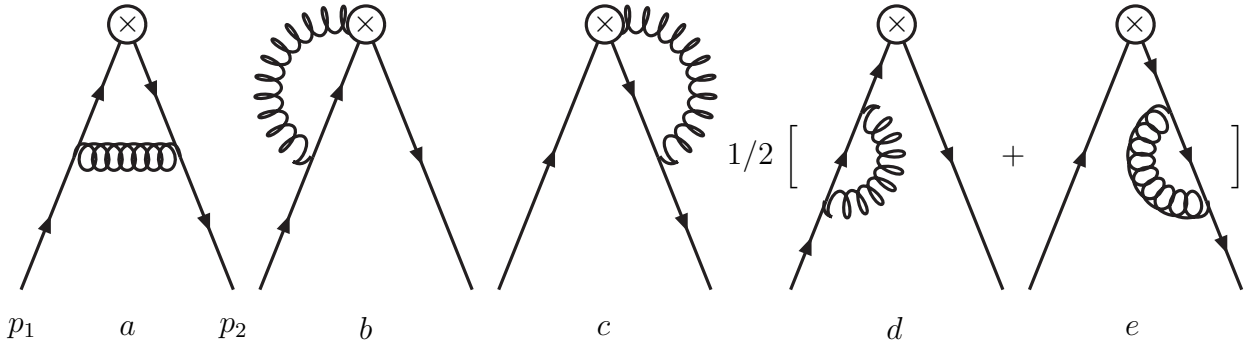
{\sf Figure~1:~The diagrams for the LO anomalous dimension for (non) 
forward scattering. {\boldmath $\otimes$} are the operator insertions
Eq.~(\ref{eq1},\ref{eq2}). The curled and straight lines denote the gluon
and quark lines, respectively.}
\normalsize

\vspace{2mm}

In the limit $D=4 - \varepsilon  \rightarrow 4$ the term $I_a$ vanishes 
in the Feynman--gauge since the contraction of the Dirac--matrices in the
numerator results into a term $\propto \varepsilon$, which cancels the 
pole of the loop--integral. The $Z$-factor, keeping the quark momentum
off-shell,  is
\begin{eqnarray}
\label{eq5}
Z_n(\mu) = 1 + 2 a_s C_F \left[\frac{2}{\ep}
- \ln\left(\frac{-p^2}{\mu^2}\right)\right] 
\left\{ 1 + 4 \sum_{j=2}^n \frac{1}{j} \right\}~,
\end{eqnarray}
with $a_s=\alpha_s/(4\pi)$, $\alpha_s = g^2/(4\pi)$, and $C_F = 4/3$. 
The first and second summand in the brackets are due to $I_b+I_c$ and
$I_d+I_e$, respectively. By Eq.~(\ref{eq3}) the anomalous dimension
\begin{eqnarray}
\label{eq6}
\gamma_n^{h_1}(g) =  a_s  C_F
\left\{ 1 + 4 \sum_{j=2}^n \frac{1}{j}\right\} = 2 a_s C_F \left\{
-\frac{3}{2} + 2 S_1(n) \right\} \equiv  - \int_0^1 dz z^{n-1} P_{h_1}(z)
\end{eqnarray}
is obtained in agreement with the result of Ref.~\cite{AM}, where
$S_1(n) = \sum_{j=1}^n (1/j)$. Note that we did not assume any sum--rule 
in this derivation. The splitting function $P_{h_1}(z)$ reads
\begin{eqnarray}
\label{eq6a}
P_{h_1}(z) = 2 a_s C_F \left\{ -2 + \frac{2}{(1-z)_+} + \frac{3}{2}
\delta(1-z)\right\}~.
\end{eqnarray}
\subsection{Non-forward-Scattering Anomalous Dimension}

\vspace{1mm}
\noindent
The diagrams in figure~1 can also be used to calculate the anomalous 
dimension of $h_1(x,Q^2)$ in the non--forward case, $p_1 \neq p_2$.
The operator insertions corresponding to Eqs.~(\ref{eq1},\ref{eq2})
read:
\begin{eqnarray}
\label{eq7}
O^{\mu} &=& -i \xt_\nu  \sigma^{\mu\nu} e^{i \xt.p_+ \kappa_+}
\left[e^{i \xt.p_- \kappa_-} -
      e^{-i \xt.p_- \kappa_-} \right] \\
\label{eq8}
O^{\mu\lambda} &=& i g t_a \xt_\nu \sigma^{\mu\nu} \xt^\lambda  
e^{i \xt.p_+ \kappa_+} \left[e^{i \xt.p_1 \kappa_-} -
e^{-i \xt.p_1 \kappa_-}\right] \frac{e^{i \xt.k \kappa_2} -
e^{i \xt.k \kappa_1}}{\xt.k}~.
\end{eqnarray}
Here we follow the notation of Ref.~\cite{BRGR}, where the corresponding
insertions for the unpolarized scalar operators were given. $\xt$ is
a light--like vector, $\lambda$ the gluon vector index and $k$ the gluon
momentum. $\kappa_1$ and $\kappa_2$ are light-cone marks with
$\kappa_{\pm} = (\kappa_2 \pm \kappa_1)/2$, and $p_{\pm} = p_2 \pm p_1$.
We work in the Feynman--gauge again and choose $\kappa_+ = 0, 
\kappa_- = 1$ in the following.

There exist various equivalent representations for the non--forward
splitting functions, which are known as $\alpha-$, $w-$, and near--forward
representation, see~[13--16]. We will first refer to the
$\alpha$--representation, which is directly related to the 
Feynman--parameter representation of the diagrams in Figure~1. Let us
denote the different contributions to the non--forward $Z-$factor, 
cf.~Ref.~\cite{BGR2},  by $J_k$,~$k=a, ..., e$.
$J_a$ vanishes for the same reason as in the forward case.
In the $\alpha$--representation the non--forward splitting function is 
obtained by
\begin{eqnarray}
\label{eq9}
K^{h_1}(\alpha_1,\alpha_2) = \frac{\alpha_s}{2\pi} C_F
\left\{\left[-\delta(\alpha_1) - \delta(\alpha_2) + \delta(\alpha_1)
       \left(\frac{1}{\alpha_2}\right)_+ + \delta(\alpha_2)
       \left(\frac{1}{\alpha_1}\right)_+ \right] 
       + \frac{3}{2} \delta(\alpha_1) \delta(\alpha_2) \right\}~.
\end{eqnarray}
The $[~~]_+$ distribution is defined by
\begin{eqnarray}
\label{eq10}
\int_0^1 dx [f(x,y)]_+ \phi(x) = \int_0^1 f(x,y) [\phi(x) - \phi(y)],
\end{eqnarray}
where $f$ is a distribution out of the space 
${\cal D}'[0,1] \times [0,1]$ and $\phi(x)~\epsilon~{\cal D}[0,1]$ a
general basic function,~\cite{VLA}.
The first addend in Eq.~(\ref{eq9})
is due to $J_b + J_c$ and the second due to
$J_d + J_e$. Eq.~(\ref{eq9}) agrees with a result in Ref.~\cite{BMLO},
see also \cite{HJ}. There are different possibilities to derive the
forward anomalous dimension from Eq.~(\ref{eq9}). One can either use
the near--forward representation and perform the limit 
$\tau = \xt.p_-/\xt.p_+ \rightarrow 0$, cf. Appendix~D in 
Ref.~\cite{BGR2}, or perform a direct integral in the $\alpha-$ or 
$w-$representation~\cite{BGR1,BGR2}, or covert $K^{h_1}$ into the local 
representation with respect of the two Mellin--variables $n$ and $n'$, 
cf.~\cite{BRGR,BGR2} and obtain the forward anomalous dimension by
setting $n=n'$.

We first change to the $w-$representation
\begin{eqnarray}
\label{eq11}
K^{h_1}(\alpha_1,\alpha_2) D\alpha =  \widetilde{K}^{h_1}(w_1,w_2) Dw,
\end{eqnarray}
with
\begin{eqnarray}
\label{eq12}
w_1 = \alpha_1 - \alpha_2,~~~~w_2 = 1 - \alpha_1 - \alpha_2~,
\end{eqnarray}
\begin{eqnarray}
\label{eq13}
\widetilde{K}^{h_1}(w_1,w_2) &=& - 2 a_s C_F \left\{ \delta(1- w_2 + w_1)
 \left[ 1 - \frac{2}{(1 - w_2 - w_1)_+} \right] \right. \nonumber\\ & &
    ~~~~~~~ \left.
+           \delta(1- w_2 - w_1)
\left[ 1 - \frac{2}{(1 - w_2 + w_1)_+} \right] - \frac{3}{2} \delta(w_1)
\delta(1-w_2) \right\}~,
\end{eqnarray}
and
\begin{eqnarray}
\label{eq14}
D \alpha &=& 
d \alpha_1 d \alpha_2 \left[\theta(\alpha_1) \theta(\alpha_2)
\theta(1-\alpha_1-\alpha_2) + \theta(1-\alpha_1) \theta(1-\alpha_2)
\theta(\alpha_1+\alpha_2-1)\right] \\
\label{eq15}
Dw &=& \frac{1}{2} dw_1 dw_2 \nonumber \\ & &\times
\left[\theta(1+w_1-w_2) \theta(1-w_1-w_2)
\theta(w_2) + \theta(1+w_1+w_2) \theta(1-w_1+w_2) \theta(-w_2)\right]
\end{eqnarray}
The forward scattering splitting function is derived by
\begin{eqnarray}
\label{eq16}
P^{h_1}(z) =  \int_{-1+z}^{1-z} d w_1 \widetilde{K}^{h_1}(w_1,z),
\end{eqnarray}
cf.~Ref.~\cite{BGR1,BGR2}. One obtains from Eq.~(\ref{eq13},\ref{eq16})
\begin{eqnarray}
\label{eq17}
P_{h_1}(z) = 2 a_s C_F \left\{ -2 + \frac{2}{(1-z)_+} + \frac{3}{2}
\delta(1-z)\right\}~.
\end{eqnarray}
in agreement with Ref.~\cite{AM}.

Likewise one may first calculate the local non--forward anomalous 
dimension $\gamma_{h_1}^{nn'}$
\begin{eqnarray}
\label{eq17a}
\gamma_{h_1}^{nn'} = 2 a_s C_F \left\{
\left[ - \frac{3}{2} + 2 S_1(n)\right] \delta_{nn'}
- \frac{2}{n-n'} \frac{n'}{n}\theta_{nn'}\right\}~,
\end{eqnarray}
where $\delta_{nn'}$ denotes the Kronecker symbol and $\theta_{nn'} = 1$
for $n' < n$ and $0$ otherwise. In the forward limit $n=n'$ 
Eq.~(\ref{eq6})
is obtained in agreement with Ref.~\cite{AM}.
\subsection{Derivation of the Anomalous Dimension from a Forward
Compton Amplitude}

\vspace{1mm}
\noindent
The anomalous dimension for $h_1(x,Q^2)$ may be also calculated form
the short distance expansion of the Compton amplitude of an axial-vector
and scalar current $j_{1}$ and $j_2$ directly using the method by Ioffe
and Khodjamirian~\cite{IK}. This has been tried in Ref.~\cite{MKS} 
recently. To extract the anomalous dimension of the corresponding 
operator matrix element in this method we start to write down the
renormalization group equation (RGE) for the Green's function
$F(q;p_1, \ldots, p_n)$, the Fourier transform of
$\langle 0|T[j_1(x) j_2(0) \phi(x_1) \ldots \phi(x_n)]|0\rangle$,
\begin{eqnarray}
\label{eq23}
\left[{\cal D} + \gamma_{j_1}(g) + \gamma_{j_2}(g) - n \gamma(g)\right]
F(q;p_1, \ldots, p_n) = 0~,
\end{eqnarray}
with $n=2$. The RG-operator is given by
\begin{eqnarray}
\label{eq24}
{\cal D} \equiv
 \mu^2 \frac{\partial}{\partial \mu^2} + \beta(g) \frac{\partial}
{\partial g} - \gamma_m(g) m \frac{\partial}{\partial m}~,
\end{eqnarray}
and $\gamma_{j_k}(g), \gamma(g), \gamma_m(g)$ are the anomalous dimensions
of the currents $j_k$, the outer legs, and the mass, respectively.
$\beta(g)$ denotes the $\beta-$function and $\mu$ is the factorization
scale. At short distances the Green's function has the representation
\begin{eqnarray}
\label{eq25}
F(q;p_1, \ldots, p_n) = \sum_k C_k(q) E^k(p_1, \ldots, p_n)~,
\end{eqnarray}
with $C_k(q)$ the Wilson coefficients and $E^k(p_1, \ldots, p_n)$ the
hadronic matrix elements. Evaluating the Compton amplitude~Figure~2 one
obtains logarithmically divergent contributions which belong to the
RGE of the Wilson coefficient, 
\begin{eqnarray}
\label{eq26}
\left[{\cal D} + \gamma_{j_1}(g) + \gamma_{j_2}(g) -  \gamma_{O_k}(g)
\right] C_k(q) = 0~.
\end{eqnarray}
The corresponding RGE for the hadronic matrix elements $E^k$ follows
from Eqs.~(\ref{eq23},\ref{eq26}).
Here $\gamma_{O_k}(g)$ is the anomalous dimension of the composite
operator $O_k(0)$ being obtained from
\begin{eqnarray}
\label{eq28}
j_1(x) j_2(0) = \sum_k C_k(x) O_k(0)~.
\end{eqnarray}
$\gamma_{O_k}(g)$ is therefore given by 
Eqs.~(\ref{eq3},\ref{eq5},\ref{eq6}). The term $\propto \ln(-p^2/\mu^2)$
which results from the diagrams of Figure~2 

\vspace*{-2.0cm}
\begin{center}
\begin{picture}(270,100)(0,0)
\setlength{\unitlength}{0.15mm}
\SetWidth{1.2}
\ArrowLine(-100,-100)(-90,-75)
\ArrowLine(-90,-75)(-80,-50)
\ArrowLine(-80,-50)(-20,-50)
\ArrowLine(-20,-50)(-10,-75)
\ArrowLine(-10,-75)(0,-100)
\Gluon(-90,-75)(-10,-75){5}{9}
\Photon(-100,0)(-80,-50){3}{5}
\DashLine(-20,-50)(0,0){10}
\setlength{\unitlength}{1pt}
\Text(-100,-115)[]{$p$}
\Text(0,-115)[]{$p$}
\Text(-100,15)[]{$q$}
\Text(0,15)[]{$q$}
\Text(-97,-50)[]{$\gamma_{\mu} \gamma_5$}
\Text(-10,-50)[]{{\boldmath $1$}}
\Text(-50,-115)[]{$a$}
\Text( 60,-115)[]{$b$}
\Text(170,-115)[]{$c$}
\Text(280,-115)[]{$d$}
\setlength{\unitlength}{0.15mm}
\ArrowLine(10,-100)(20,-75)
\ArrowLine(20,-75)(30,-50)
\ArrowLine(30,-50)(60,-50)
\ArrowLine(60,-50)(90,-50)
\ArrowLine(90,-50)(110,-100)
\GlueArc(30,-50)(26.92582,-113.578,0){5}{8}
\Photon(10,0)(30,-50){3}{5}
\DashLine(90,-50)(110,0){10}
\ArrowLine(120,-100)(140,-50)
\ArrowLine(140,-50)(170,-50)
\ArrowLine(170,-50)(200,-50)
\ArrowLine(200,-50)(210,-75)
\ArrowLine(210,-75)(220,-100)
\GlueArc(200,-50)(26.92582,180.0,293.5782){5}{8}
\Photon(120,0)(140,-50){3}{5}
\DashLine(200,-50)(220,0){10}
\ArrowLine(230,-100)(250,-50)
\ArrowLine(250,-50)(270,-50)
\ArrowLine(270,-50)(290,-50)
\ArrowLine(290,-50)(310,-50)
\ArrowLine(310,-50)(330,-100)
\GlueArc(280,-50)(10,180.0,0){5}{8}
\Photon(230,0)(250,-50){3}{5}
\DashLine(310,-50)(330,0){10}
\setlength{\unitlength}{1pt}
%
\end{picture}
\end{center}

\vspace*{4cm} \noindent
\small
\begin{center}
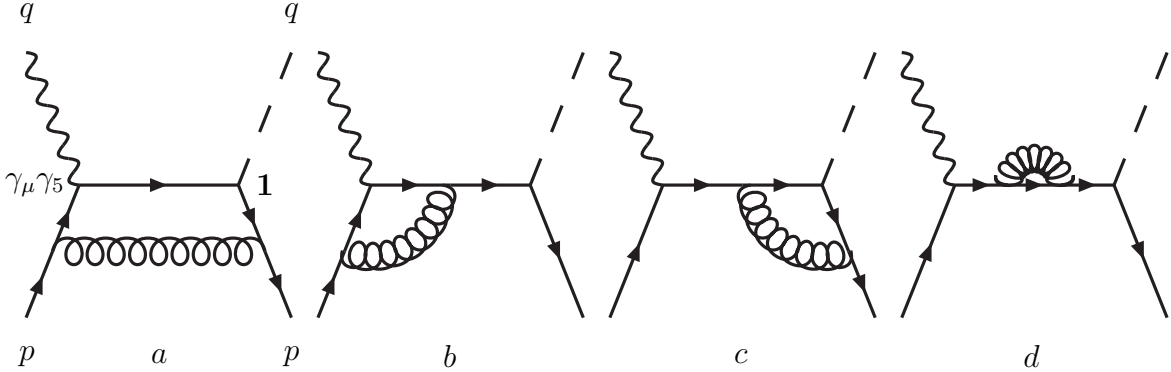
{\sf Figure~2:~QCD 1--loop diagrams to the forward Compton amplitude.}
\end{center}
\normalsize
%
is
\begin{eqnarray}
\label{eq29}
\gamma_C(g) = \gamma_{j_1}(g) + \gamma_{j_2}(g) - \gamma_{O_k}(g)
= - 2 a_s C_F \cdot 2 S_1(n)~.
\end{eqnarray}
The anomalous dimensions of the two currents are~\cite{GRAC}
\begin{eqnarray}
\label{eq18}
\gamma_{j(s)} &=& -C_F a_s (s-1) \left\{ (s-3) + \frac{a_s}{18} \left[
4 (s-15) T_F N_f + (18 s^3 -126 s^2+163 s +291) C_A \right. \right.
\nonumber\\  &  &~~~~~~ \left. \left.
- 9 (s-3) (5 s^2 -20 s
+1) C_F \right] \right\} + O(a^3_s)~,
\end{eqnarray}
where $s$ denotes the spin, i.e.  $\gamma_{j(s=1)} \equiv 0$ and
\begin{eqnarray}
\label{eq19}
\gamma_{j(s=0)} = -3 C_F a_s 
+ C_F \left[\frac{10}{3} T_F N_f - \frac{97}{6}
C_A - \frac{3}{2} C_F \right] a_s^2 + O(a^3_s)~,
\end{eqnarray}
which yields the splitting function, Eq.~(\ref{eq8}). This shows that the
method of Ref.~\cite{IK} can be used to calculate the anomalous 
dimensions.
\section{
A Remark on the  Anomalous Dimension of {\boldmath $h_1(x,Q^2)$} at NLO}

\vspace{1mm}
\noindent
The twist--2 NLO anomalous dimension and splitting function for the 
distribution $h_1(x,Q^2)$ was calculated in Refs.~\cite{NLO} using the 
cut vertex method of \cite{CUT} or the method outlined in \cite{CFP}.
In both cases the $\delta(1-z)$--contribution cannot be fixed easily by a 
direct calculation, see also~\cite{DEL}, since the calculation is 
performed in a physical gauge. In fact, a complete calculation of these 
terms is not yet available. On the other hand, complete calculations of 
the NLO anomalous dimensions as occurring for the various polarized and 
unpolarized non--singlet and singlet twist--2 parton densities 
contributing to the structure functions exist in the $R_\xi$--gauges, 
cf.~Refs.~\cite{COV}. Thus these direct calculations assure that for the 
non--singlet $`-'$ twist--2 quark splitting functions fermion number
conservation holds, as energy--momentum conservation holds in the singlet
sector.

Knowing this result and working in a class of gauges in which the NLO 
anomalous dimensions are gauge invariant one may take advantage in 
comparing the $\delta(1-z)$--terms occurring in the quark splitting 
functions $P_{NS^-}^q(z)$ and $P_{h_1}(z)$. This is now done in the axial 
gauge. One classifies the contributing Schwarz {\it distributions} dealing 
with the $+$--distribution as an individual one and {\it not} being 
expressed
due to  a composed `function' of $\delta-$ and Heaviside functions in
the limit $\varepsilon \rightarrow 0^+$. Then all the $\delta(1-z)$ terms
arise from the self-energy terms, which are {\it independent} of the
operator insertion. Due to this they can be determined without an
explicit calculation, completing the calculation of the splitting 
function. The result on the forward scattering splitting functions has 
been extended to the case of non--forward scattering in \cite{BMNLO}.
\section{Summary}

\vspace{1mm}
\noindent
We have shown by three different complete calculations that the 
calculation of the leading order anomalous dimension of the transversity 
distribution $h_1(x,Q^2)$ as originally being derived in Ref.~\cite{AM}
is correct. Our results also agree with those which have been obtained in
Ref.~\cite{MC} very recently. It was shown that the method of Ioffe and 
Khodjamirian can be used to derive anomalous dimensions in the present
example and similarly for related cases if applied to coefficient
functions $C_k(x)$, contrary to the result of \cite{MKS}, which has to 
be regarded to be wrong. The methor of Ref.~\cite{MC} may thus be
of importance for higher order calculations in situations in which the
light cone expansion does not apply. Furthermore we argued that the
$\delta(1-z)$ terms of the next-to-leading order calculations, although
not yet being calculated directly, are right.

\vspace{1mm}
\noindent
{\bf Acknowledgement}.~For a conversation I would like to thank W.L. van
Neerven and D.~Robaschik. My thanks are due to D. Boer, B. Pire, and
R. Kirschner and W. Vogelsang for their interest.

\vspace{1mm}
\noindent
{\sf Note added in proof.}~After this paper was finished D. Boer
pointed out to me that the anomalous dimension of $h_1(x,Q^2)$ was 
calculated prior to Artru and Mekhfi \cite{AM} by Shifman and Vysotsky
\cite{SV} and at the same time by Baldraccini et al. \cite{BCRS}. Later
also Bukhvostov et al. calculated a Matrix element in the quasipartonic
approximation which can be interpreted as the LO anomalous dimension
\cite{BUKH}. I would like to thank R. Kirschner for this remark.
We also note that the anomalous dimension of $h_1(x,Q^2)$ at small $x$
was calculated in Ref.~\cite{KIRS}.

\end{document}